\documentclass[prd,a4paper,superscriptaddress,aps,showpacs,floats,nofootinbib,linenumbers,twocolumn]{revtex4-1}
\usepackage{epsfig}
\usepackage{graphicx}
\usepackage{amsmath}
\usepackage{amssymb}
\usepackage{color}
\usepackage{hyperref}
\usepackage{url}
\usepackage{float}
\usepackage{placeins}

\let\Oldsection\section
\renewcommand{\section}{\FloatBarrier\Oldsection}

\let\Oldsubsection\subsection
\renewcommand{\subsection}{\FloatBarrier\Oldsubsection}

\let\Oldsubsubsection\subsubsection
\renewcommand{\subsubsection}{\FloatBarrier\Oldsubsubsection}
\usepackage{subcaption}
\usepackage{setspace}

\makeatletter
\gdef\@ptsize{2} 
\makeatother

\begin{document}

\nolinenumbers
\title{IMDb data from Two Generations, from 1979 to 2019; Part one, Dataset Introduction and Preliminary Analysis}

\author{M. Bahraminasr}
\email{majid.bahrami.nasr@gmail.com}

\author{ A. Vafaei-Sadr}
\email{vafaei.sadr@gmail.com}

\affiliation{Institute for research in fundamental sciences (IPM), PO Box 19395-5531, Tehran, Iran}

\begin{abstract}
``IMDb.com'' as a user-regulating and one the most-visited portal has provided an opportunity to create an enormous database. Analysis of the information on Internet Movie Database - IMDb, either those related to the movie or provided by users, would help to reveal the determinative factors in the route of success for each movie. As the lack of a comprehensive dataset was felt, we determined to do create a compendious dataset for the later analysis using the statistical methods and machine learning models; It comprises of various information provided on IMDb such as rating data, genre, cast and crew, MPAA rating certificate, parental guide details, related movie information, posters, etc, for over 79k titles which is the largest dataset by this date. The present paper is the first paper in a series of papers aiming the mentioned goals, by a description of the created dataset and a preliminary analysis including some trend in data, demographic analysis of IMDb scores and their relation of genre MPAA rating certificate has been investigated.

	{\bf Keywords}:  IMDb data, statistical analysis, data science 
\end{abstract}

\maketitle


\section{Introduction}
Since the Web 2.0 era the internet usages has been revolutionized, more people accessing with higher bandwidth make the user from almost mere observers to creators and participants. Excluding private cloud storage services, the platforms created and/or owned by giants companies such as Google, Twitter, Facebook, Amazon, etc has created a gigantic data warehouse. To make this data usable for analysis the extracting, processing and organizing the is the very first and essential step. 

Internet Movie Database (IMDb) is an online database dedicated to all kinds of information about a wide range of motion picture contents such as films, TV and online-streaming shows, series, etc. The information which is presented on the IMDb portal includes cast, production crew, personal biographies, plot summaries, trivia, ratings, and fan and critical reviews and much other similar information which are mostly provided by volunteer contributors. To contribute,  registration is required, however since no legal document is required, one is able to use an arbitrary name. Being a user-regulated website could be a shortcoming as it would be vulnerable to malicious attempts from a certain group to bias information. However, taking advantage of a large community not only overpower these attempts but also create a cornucopia of valuable data that analyzing them may shed light on many hidden factors that help movie industries and other related businesses in content production. 

There are various studies have been done on IMDb data. Oghina and et al. investigated the possibility of prediction of IMDb rating using social media contents such as tweets and YouTube comments \cite{i1}.  Otterbacher showed there is a tangible difference between men and women's review writing style using the IMDb review section \cite{i2}. In \cite{i3} the connection between user voting data and economical characteristics of films such as budget and box office data has been investigated by Wasserman et al. Hsu et at., using linear combination, multiple
linear regression, neural networks predicted the IMDb rating from other movie's attributes using 32968 titles \cite{i4}; and In \cite{i5} Nithin et al. used Logistic Regression, SVM Regression and Linear Regression to predict box office data.  In \cite{i6}, using available demographic information on IMDb Bae et al. created a demographic movie recommender system. Ramos et al. showed the distribution of votes showed a scale-free behavior \cite{i7}. 

There are various datasets available each with a different policy.
IMDb itself discloses a \href{https://datasets.imdbws.com/}{subsets their data}  for personal use. Furthermore, there are more dataset available freely on kaggle such as \href{https://www.kaggle.com/stefanoleone992/imdb-extensive-dataset}{IMDb movies extensive dataset}  \href{https://www.kaggle.com/lakshmi25npathi/imdb-dataset-of-50k-movie-reviews}{IMDB Dataset of 50K Movie Reviews} , Also there are other which are required corresponding with the owner.

Seeing that many datasets available online usually do not cover some important information or they are not large enough, we determined to create a dataset that covers some drawbacks that exist in the available sets. Still, the other datasets could be used as a complement. The present paper is the first paper in a series of papers aiming to create a suitable dataset, analyze it, and predict some information using those data. 

\section{Available Data}
The created dataset is based on the data available on IMDb website and some third-party datasets and resources to provide some additional information on the available data on IMDb, such as similarity of countries and languages or how much a certain actor is talked about comparing other using the number of google results. The data mainly extracted from \href{https://imdb.com/}{IMDb Portal}, \href{https://www.indexmundi.com/}{IndexMundi}   
,\href{http://www.elinguistics.net/Compare_Languages.aspx}{Elinguistics}, and Google results in a specific field of data.  This section is dedicated to the description of gathered data from the IMDb database. The full description of the data is available at \url{https://help.imdb.com}. To learn about the gathered and processed data from IndexMundi and Elinguistics you may refer to Appendix  \ref{App1}.

To access each title, we used the code which IMDb assigned uniquely to each title. The code started with a  double t -``tt''- followed by some numbers, for example this code for the title {\it Logan (2017)} is ``tt3315342''. Using this code one can have access to the title's main page, for example the address for the title {\it Logan (2017)} would be like \url{https://www.imdb.com/title/tt3315342/}. The main portion of extracted data is from the title's page and some relative addresses from that page, for instance, the rating data extracted from the relative address of \url{/ratings} of each title page ,e.g. \url{https://www.imdb.com/title/tt3315342/ratings} for the title {\it Logan (2017)}.
\subsection{Movie Name}
The Movie name is the name which was given to each title by the producer, we found a few minor discrepancies on the titles from different part of IMDb. Here our reference is the name on the designated page for each title.
\subsection{Poster}
There are several posters associated with each title. Here the main poster which has been presented on the title's page, is stored. 
\subsection{Alternate titles (AKAs)}
Alongside the original title, every film may have other titles or names that are known with, either in different countries and/or languages; in this case, alternate titles may be listed. Default alternate title is the same as the primary one \cite{b1}. The alternate titles could be a small deviation from the original name and/or be in other languages rather than the film's language[s]; for example for the movie \href{https://www.imdb.com/title/tt3315342}{Logan (2017)} the alternate titles are mainly are the original title plus {\it Wolverine} which is sometimes in different languages rather than its original language, English. In this case, the regular NLP analysis may not give any insightful results, however, the number of the alternated title could be an interesting factor. It could somehow show how much people and/or different nations care to give the movie their own names.

\subsection{IMDb ratings and Number of votes}
Every user can vote from 1 to 10 to rate each title, there is no need to writing a review upon giving the score. A weighted average of the registered users will be shown as the title's rate.  IMDb's intention is to reduce the intended attempts to change the title rating from actual worth. Various filters are applied for this propose and IMDb does not disclose the math \cite{b2}. However the arithmetic mean is also available in the relative address of \url{/ratings} for each title. Moreover, the voting distribution histogram and demographic information of rating and number of votes are also available. Here demographic information contains the top 1,000 voters information, US and none-US users, and different age and genders. The top 1,000 voters are the top 1,000 who have voted the most titles and are unknown \cite{b3}. For the rating section, the IMDb's rating, the arithmetic mean of rating, median, and all the demographic information about the rating (by age, sex, and information on top 1000 users, US and Non-US users) and the number of votes have been gathered.

\subsection{Metacritic Score and User/Critics reviews}
Besides the rating, the \href{https://www.metacritic.com/about-metascores}{metacritic score} and user and professional critics reviews are available, so one could be informed of other viewers' opinions \cite{b2}. At first glance, the semantic analysis of each review seems to be the only way to use this information. However, the number of reviews could be a helpful factor to validate the user's ratings. Despite the reviews could be biased, ignoring various downfalls of the title, especially the one written by users rather than a renown critic, the number of them could be showing how much the title worth to people dedicate their time to write about, after watching the movie. On the other side, the votes could be blind votes which are given by particular groups very high or low, without watching the movie as it happened for \href{https://www.imdb.com/title/tt2788710/}{The interview (2014)} which at the beginning of its release get a near-perfect score \cite{b5}. Not only blind voting causes a problem, but also die-hard fans of some genres like Sci-Fi, ignoring major flaws, could also have very biased voting, However, after a given period of time the effect this attempt will smooth out. On the other hand,  writing a review is less impulsive action and needs more contemplation, and of course being a fan of a genre won't be enough to write the reason why an individual liked/disliked a title.    

\subsection{ Popularity and change }
The popularity ranking on a title separately compares movie titles with each other
\cite{b6}. Here the popularity and its changes at the time of extraction have been stored.

\subsection{Motion Picture Rating, IMDb Certificates}
To specify the appropriate audience for each title IMDb provides the   Motion Picture Rating (MPAA) certificate. Explanations for the available entries are could be found at \cite{b7}.
Each country has its own MPAA system and/or age restriction for each title. Here the rating certificate given to each title within the United States has been considered as the reference.  The information about other countries also extracted from relative url of \url{/parentalguide} for each title.

\subsection{Parental Guide}
IMDb includes parental guide entry to provide the parents with additional information by describing some scenes to determine the appropriateness of each title\cite{b8}.
All the information is available in the relative address of \url{/parentalguide} of each title. The entries include {\it Sex and Nudity}, {\it Violence and Gore}, {\it Profanity}, {Alcohol, Drugs, and Smoking}, and {\it Frightening, and Intense Scenes}. Here just the number of scenes ( and not the description) and, if it was available, the degree of severity (Mild, Moderate, Severe) are extracted.

\subsection{Genres}
There are several genres, which each title may associate with one and more. For the full description you may refer to \cite{b9}.

\subsection{Countries and Languages}
Country is defined as the country where the production company is based. It is possible multiple companies are associated with each title \cite{b10}.
The languages which are spoken in each title are listed in order of frequency \cite{b11}.

\subsection{Release Dates and Locations, Filming Dates and Locations  }

Release dates and locations have been gathered from the relative address  \url{/releaseinfo} of each title. This portion of data could be an indicator of the potential popularity. For example, if the title released in different countries in a small time window it may be a sign for its popularity. 

Moreover, the filming dates and locations have been extracted from relative address of \url{/locations} . The filming locations could be a good indicator for the budget class of the movie especially when no data is available on the budget.

\subsection{Box Office data - may need to add}
The extracted data here are: Budget, Opening Weekend USA Income, Opening Weekend USA, Gross USA, Cumulative Worldwide Gross.

\subsection{ Director, Writers, Stars }
Director, writers, stars, and roles are also extracted. There is an elaborate list for each of them available but at this point, for the sake of simplicity, the first names on the main page of each title are stored. To machine they are some random string. Plus, there are not a lot of data to assign them a value or a vector with techniques such as  Word2Vec. Some datasets are containing the number of Facebook page's likes for each actor or similar information like \href{https://www.kaggle.com/carolzhangdc/predict-imdb-score-with-data-mining-algorithms}{this dataset on kaggle}. However the size of these datasets is limited and does not cover all the names that are needed here. Here we have taken another strategy and used the number of google results. To avoid name similarity we used the profession alongside the name to narrow down the results as much as possible; for example we searched {\it Tom Hanks + movie star}, or {\it Steven Spielberg + director}.

\subsection{Production Companies}
The list of production companies has been extracted from the relative address of \url{/companycredits} of each title.

\subsection{Related movies}
Up to twelve similar titles are suggested under the {\it ``More like this''} entry. These titles are generated based on various information such as genres, country, stars, etc \cite{b13}.
Here we also extracted the IMDb rating, number of votes, and the IMDb code for each related title.

\subsection{Keywords and Storyline}
There are also storyline plot and keywords available. This data is valuable to this extent that reveals the key and unique elements which are presented in the movie. The keywords are offered by users and they can vote if they are relevant or not. Here we gather all the keywords sorted by a relevancy score from the relative address of \url{/keywords} which is calculated by this relation:

\[\text{Number of votes}\times\frac{\text{Number of positive votes}}{\text{Number of votes}}\]

\section{Data Cleaning and Processing}
Here we briefly describe the pre-processes and labeling format that is essential to know before using the data.

\subsection{Structure of the Data}   
\subsubsection{Data Format}
Data is packed according to the release year of each title for better management. All the data are stored in a CSV file with UTF-8 encoding.
The index of the table has been set to its unique IMDb code. Using the IMDb code as the index could be beneficial during the model training since it uniquely determines the title it does not contain specific information that could be used during the analysis or model training to be a part of the table.
Moreover, there is a subdirectory for each year containing the film's poster in jpg format each with the dimension of 182$\times$268, 72 DPI. The size of the data is around 5-25 Mb for the CSV file and 10-15 kb for each  poster image. There are 79793 rows of data, and 67393 poster files are available in total.
\subsubsection{Columns' names}
Since heavily relying on column numbers in the middle of analysis could be confusing, especially here which data are packed according to the release date of the titles and the number of columns may vary.  Consequently, we introduce a specific wildcard access data columns. Including those patterns enable the users to search with Regular Expressions (RegEx) to narrow down the list of columns to the specific part of the table. 
Here we used capital letters at the end of each column name to distinguish them from the actual name of each column; since the multi-parted names are accompanied by underscore, using python regular expression has been made easy. 

\subsection{Wildcards}
Here we will briefly describe the wildcards' meaning 
\begin{itemize}
\item{\texttt{*\_GS}} GS stands for General Set, which contains general information about the title such as the name and alternate names, technical information like runtime sound mixing, the plot, keywords, related movies, filming locations and companies, etc.

\item{\texttt{*\_GENRE}}  This wild card is related to information about the genre. Since Each title's genre does not necessarily fall into one category, here we created two sub-wildcard of {\texttt{*\_SET\_GENRE}} for a complete set of genre and  {\texttt{*\_HOTVECTOR\_GENRE}} for their hot vector representation

\item{\texttt{*\_COUNTRY}} With this wildcard you may access the country information of each title. There are two sub-wildcards are also available  \texttt{*\_SET\_COUNTRY} \texttt{*\_HOTVECTOR\_COUNTRY} for list of country and hot vectors of countries respectively. Here we included two quantized information about the country; the reference of comparison has been chosen the United States as the creator of the most titles each year. In \texttt{*\_NONGEO\_DIS\_COUNTRY} the mean Manhattan distance between 106 parameters has been calculated, for more information about this analysis please refer to Appendix \ref{App1}.
\texttt{*\_GEO\_DIS\_COUNTRY} provides information about the geographical distance by calculating the great-circle distance between the country's capital from Washinton DC using haversine formula.

\item{\texttt{*\_LANGUAGE}} This wildcard related to languages which are spoken in the original version of each title. \texttt{*\_SET\_LANGUAGE} includes list of spoken languages with descending order of usage frequency. \texttt{*\_HOTVECTOR\_LANGUAGE} is the hot vector of languages. Language comparison to English is stored in \texttt{*\_ENGLSIH\_DIFF\_LANGUAGE} column. {\texttt{*\_GOOGLE\_RES\_LANGUAGE}} contains the number of google search results. It is abundantly clear that the exact number is not a good reference but its order of magnitude would give an idea of how much a language is spoken about relative to another. Although the number of people who are speaking a certain language as the first and/or second language also might be a good option to assign a meaningful value to each language, however, we hadn't found any resource for all the languages.

\item{\texttt{*\_BOXOFFICE}}  This wildcard is the data related to Boxoffice, Please note that the Currency is not converted to their today's value.

\item{\texttt{*\_DWSC}} This wildcard is related to Directors,  Writes, Stars and their roles, and Production Companies. There is a comprehensive list for each field but here the list of names is restrained to the names which are appeared on the main title page. Also the number of google results for directors, writers and stars are included in sub-wildcard of \texttt{*\_GOOGLE\_RES\_DWS}

\item{\texttt{*\_RATING}} \texttt{*\_RATING} is associated with the voting, the rate and the number of votes. The general information such as the total number of votes, arithmetic mean rating, and IMDb rating and median of votes can be found using \texttt{*\_G\_RATING} wildcard. The sub-wildcard related to the distribution of voting  are 
\texttt{*\_NUM\_DIST\_RATING}, \texttt{*\_PERCENT\_DIST\_RATING}, which are assigned to the number of specific vote and the percentage respectively.
US and Non US voters, sore and number of vote are accessible using the wildcards of\texttt{*\_SCORE\_GIS\_RATING}, and \texttt{*\_NUM\_GIS\_RATING}. Top users score and number of votes are in
\texttt{Top\_1000\_Voters\_SCORE\_DEMOGRAPHIC\_TOP\_RATING}
, \texttt{Top\_1000\_Voters\_NUM\_DEMOGRAPHIC\_TOP\_RATING} columns.
For all Ages and gender and/or separately sorted by age intervals and gender the wildcards of
\texttt{*\_SCORE\_DEMOGRAPHIC\_AG\_RATING},
\texttt{*\_NUM\_DEMOGRAPHIC\_AG\_RATING} are used to access the score and number of votes respectively.

\end{itemize}
 
\subsection{Data access}
To access the data you may contact us. Moreover, to have a glimpse of how data looks like, a portion of data is available \href{https://github.com/mjdbahram/IMDb-sample-data}{at this repository}. 

\section{Preliminary Analysis and Discussion}
This section aims to demonstrate an overview of data. Here we are going to study some trends from 1979 to 2019. Moreover, we are going to study the effect of other factors such as genre and parental guide information on the IMDb rating, distribution of ratings, and demographic information of ratings.  

\subsection{Trends }
There are two types of information available on IMDb; one of them is related to attributes of each title such as the runtime, genre, etc, another is created by users' activities such as voting. An important point is that the first type could be assigned to the release year but the latter could not. The votes, For instance, could be cast in any year so speaking about the trends on this portion data should be interpreted as the scores that are given to the title released on a specific year, not the scores that are given within that year.

\subsubsection{Number of titles }
In this study, the titles with a vote number larger than 100 have been considered. The number of titles with this condition grows every year. However, it drops after a peak in 2017, by around 300 and 1000 numbers for 2018 and 2019 respectively, Fig \ref{vt}. Although it might be counter-intuitive, it could be an indicator that for the threshold of 100 vote numbers it takes at least three years for movies to follow the expected trend. The extrapolated values for 2018 and 2019 are 5007 and 5295 respectively.

The average of the number of votes is ascending till 2010, and drop by 7000 numbers in 2019 (Fig \ref{f3}). This is also an indicator that vote numbers need a long time to follow the expected trends. The extrapolated value for 2019 is 18182; however, the changes seem to be more drastic than the number of titles. For example, the extrapolation polynomial prediction value for the year 1994 is 10528 which is lower than its actual value, 15122. The Fig  \ref{f3} also demostrates that the males' mean vote numbers are greater than the females'. Most of the votes belong to the category of males between 30-44  and female voters under 18 are the smallest category. There are no significant changes in trends except for the category of 18-29 males, overpassed the males over 45 in 1998. The under 18s are the minor portion of voters, which sound reasonable because of the restrictions for the title they can watch and using the internet.

Another interesting trend in the data is the increasing difference between arithmetic mean and IMDb rating by 0.16 score. Moreover, the difference between male and female average scores has had 0.24 increment from 1979 to 2019. The variance of votes casted by different age category has a decreasing trend until 2009 and after that, it has an increasing trend, Fig \ref{diff}.

\subsubsection{IMDb Scores}
The scores are given by females are higher than males Fig \ref{f3}; the average of scores given by men is 5.93 but women is 6.13. Their trends are different; despite the persistent decreasing trend of men's mean scores, women's votes after a descending trend has increased in from 6.02 in 2009 to 6.20 in 2019. This could be a sign for emerging of more politically correct content and increasing the role of women in the movie industries.
The highest scores are given by males under 18 and the lowest scores are given by males over 45 years old. The score given by female voters showed fewer changes in different age categories.

\subsubsection{Top three Languages}
The languages: English, French, Hindi, Italian, Japanese, Spanish show up as the top three languages from 1979 to 2019. English is always the most frequent and its number growing each year. However, its percentage decreased from 35.4\% in 1979 to 18.6\% in 2019. The other aforementioned languages ranked as second and third interchangeably, except for the Italian which after 1995 has not been among the top three most frequent languages.

The top three most frequent countries during 1979-2019 were: France, Germany, India, Italy, UK, USA. The USA always ranked first with increasing the number of released titles each year. However, its share decreased from 30.1 to 17.5 percent. The second and third place are received by other countries interchangeably. The only significant trend dropping the number of titles produced by Italy from 1988, and the jump of Germany' after 1989.

\subsubsection{Genre and IMDb Certificates}
Drama and Comedy were two first genres and Thriller, Horror, Romance, and Action are received the third place interchangeably in different years. Each genre has specific behavior which could be caused by many reasons such as a popular actors or directors or popular stories. There are some distinguished trends like increasing the percentage of documentary and short films overtime and descending trend of fantasy genre since 1994.

The MPAA certificate of R was the most frequent. before 2014, PG-13, and after that, TV 14 comes as second and the third place is received by PG.

\subsection{Analysis of Movie Rating}
One of the most important features in the IMDb database is the IMDb rating. Moreover, distribution and demographic structure of data divided by age and gender and location of voters, and being among the top 1000 voters are available. Here we are going to study IMDb rating alongside other parameters such as genre, rating certificate, and other parental guide information.

\subsubsection{Demographic analysic of IMDb rating scores }
\label{demo_ratings}
Most of the votes, in descending order of vote numbers, are from the age category of 30-44, 18-29, over 45, and under 18. Each genre receives the highest rating, in descending order of scores, from males under  18, 18-29, 30-44, and over 45 and females 18-29, under 18, 30-44, and over 45.

The distribution of voting relative to age and sex is demonstrated in the Fig \ref{vp} . As it can be inferred from Fig \ref{vp} the female voters are slightly prone to submit higher scores than males. Moving from younger age interval to older, the tendency of giving a very high or very low score for both genders decreases. However, females' changes are slightly less than the males'. 

Fig \ref{sh} encapsulates the information about the correlations of scores for two age intervals (first row of each cell) and their respective percentage in the total population (second row of each cell); the least correlation for scores is between over 45 years old males and under 18 females and the most correlation is between females between 18-29 and 30-44. The difference between the maximum and minimum of correlations between men is 0.09 and between women is 0.15. An interesting point between All these data here is that either we look at the auto-correlation of each gender or cross-correlation of males and females, the most corrected part of each block belongs to age categories of 18-29 and 30-44  Which according to the second row of each cell in Fig \ref{sh}, create the largest portion of the voters' populations.

IMDb rating tends to be more correlated with Top 1000 users score than Arithmetic mean.
In both IMDb rating and Arithmetic mean of votes are more correlated with Non-US's scores than US user, however, correlation of IMDb ratings has become more correlated with the Non-US voters than US voters

Moreover, IMDb rating is more correlated with males' scores than females' scores. In males category, IMDb rating becomes more correlated than Arimethic mean by moving to the older age categories however in females the increments are smaller and the most changes belong to females between 30-44 years old. For both genders the effect of under 18-year-old voter are decreased in IMDb rating, Fig\ref{rating-AAM}.

\subsubsection{Genres and IMDb rating scores }
The three most highly scored genres in IMDb are Drama, Comedy, and Action with the score means of 5.96, 5.91, 5.88; and the least scores are given to Musical, Western, and Sci-Fi with score means of 5.78, 5.78, 5.77. In each MPAA certificate, the most highly scored genres are almost the same, Drama and comedy are the most frequent genre in each certificate.

The distribution of votes for each genre reveals a lot about the fans. The most frequent scores are 10, 7, and 6 without any exceptions. Mostly they receive 7 and after that 6. However, for the genre like Sci-Fi and Western and Musical, the score 10 is the most frequent.

 As males are constitute the major portion of voters the demographic gender-neutral data follows males data.
Males' number of votes obey the same behavior similar to average behavior mentioned in the section \ref{demo_ratings} for all the genre except for the Musical, and Western, males over 45 years old votes the most after the 30-44 category. The distribution of the scores is like the overall age distribution. However, females' average scores show dependence on the genre. The order of given scores, with descending order of mean scores, are as follows: Action, Adventure, and drama: 18-29, under 18, over 45, 30-44 years old. Animation, Biography, Comedy, Crime, Documentary, Family, Fantasy, Mystery, Romance, and Thriller: 18-29, over 45, under 18, 30-44 years old. History, Horror, Musical, Sci-Fi, Short, Sport, War, and Western: 18-29, over 45, 30-44, under 18. It is worth mentioning that all the order of the mean number of votes distribution is similar to the gender-neutral age categories.

\subsubsection{MPAA rating and Paretal guide informa IMDb rating scores }
Movies with general audience receive the highest mean IMDb rating, 6.39; and the certificates which require the minimum age of 18 years old have a minimum score of 5.6.

Correlation between number scene containing Sex and Nudity, Violence and Gore, Profanity, Alcohol Drugs and Smoking, Frightening and Intense Scenes, Scenes show almost no correlation with the IMDb scores Fig \ref{f3} (numbers are multiplied by 100). However, males scores are slightly more correlated.

\section{Conclusion}
In this paper, we introduced the largest and the most comprehensive movie database created based on the IMDb dataset. The database contains a variety of information of over 79k titles, ranged from alternate titles, genres, MPAA rating certificates, and related movies information to demographic information on IMDb ratings. Other unique features that make this database special in deep learning and machine learning model training is that we tried to make it as quantize as possible. Alongside the name of countries, languages, actors and actresses, directors, writers, and movie companies, we assigned a number with methods that have been explained in the appendix \ref{App1}. Moreover, the main poster of each title has been included.

A preliminary analysis of the data has been presented. There were some interesting trends, for example, the difference between IMDb rating and Arithmetic mean of scores and also the difference between male and female mean scores have been increased over 1979 to 2019; in this time interval, the males' votes constitutes the major portion of voters' population. Males' votes average had a decreasing trend while females' started to increase after 2009; moreover, females' give a higher score than males'. 

Also, the data demonstrated that the country Italy and Italian language have not been ranked in the top three languages and countries since 1988 and 1995 respectively.

The most frequent genres were Romance and Drama and the most frequent MPAA rating certificates were R.

Analysis demographic information of IMDb rating revealed that most voters are in the age category of 30-44; and after that, the categories of 18-29, over 45 and under 18 years old are, respectively, the most voters. The score given by males in each category has an inverse relation by the age. Females also follow this pattern except the highest score comes from the category of 18-29 years old rather than under 18 categories. Female votes in different age categories depend on the genre but males showed a more consistent pattern.

The most frequent votes for each genre are 7 and 6 but in some genres like sci-Fi, Western, and Musical the most frequent vote is 10.

Study of correlation of demographic information of scores revealed that ages of 18-29 and 30-45 are the most correlated ones, either in the correlation between each gender or in the correlation of males and females. Two other age intervals of under 18 and over 45 show to be more inconsistent.

The number of scenes which include Sex \& nudity, Violence \& Gore, Profanity, etc has an almost-zero correlation with the IMDb rating.

The IMDb rating is calculated somehow that has become more correlated with Non-US voters,  Top 1000 voters, and male voters.

\appendix
\subsection{Third-party data compliments}\label{App1}
Since part of data is in form of text, we needed to utilize an appropriate approach to turn them into numbers so we can use it in training of machine learning models. Despite turning them into hot-vectors might sound like the only option, we used other approaches to assign each entry a suitable value. Here we briefly discuss about the datasets and the process of preparation. As the results of some policy we are not allow to reshare some of these third-party data, thus only our results after the processing will be disclosed. 
\subsubsection{IndexMundi}
All the data are available \url{https://www.indexmundi.com/factbook/compare}. Please read carefully \href{https://www.indexmundi.com/help/terms/}{the Term of use} before using their data. Here we mainly used demographics information, and some information from geography and economy table. In total, 106 Fields of data extracted. All fields of data are normalized to so they are ranged from 0-1. Since we need to assign each country a value we calculate the geographical distance and non-geographical distance using extracted data from United States. The missing information was another issue; Antarctica, for instance, does not possess 96 out of 106 our data columns. Here we take the availability of data as similarity factor, therefore the number of missing data will increase the distance of two country. Here, we report the mean Manhattan distance as the non-geographical distance of countries. In this process the most similar, excluding geographical distance, countries was United Kingdom and the least similar was Antarctica, which sounds reasonable.
\subsection{Elinguistics}
This database used to compare different languages to English. Despite their similarity to English, the most spoken language, could be consider as an important factor, this analogy could be misleading since two different languages from English might be highly similar. The reported values are from 1 to 100.  Highly related languages, Related languages, Remotely related languages, Very remotely related languages, and No recognizable relationship receive score Between 1 and 30, Between 31 and 50, Between 51 and 70, Between 71 and 78, and Between 71 and 100 respectively. You may learn more about their analogy from \href{http://www.elinguistics.net/Language_Evolution.html}{their methodology}.
\subsection{Number of Google Results}
The number of google results also used to compare how much the searched key is talked about on the web. For the languages, stars, directors, writers, and production companies, we used number of google results.

\begin{widetext}
\begin{center}
	\begin{figure}[H]
			\centering
		\includegraphics[width=0.7\textwidth]{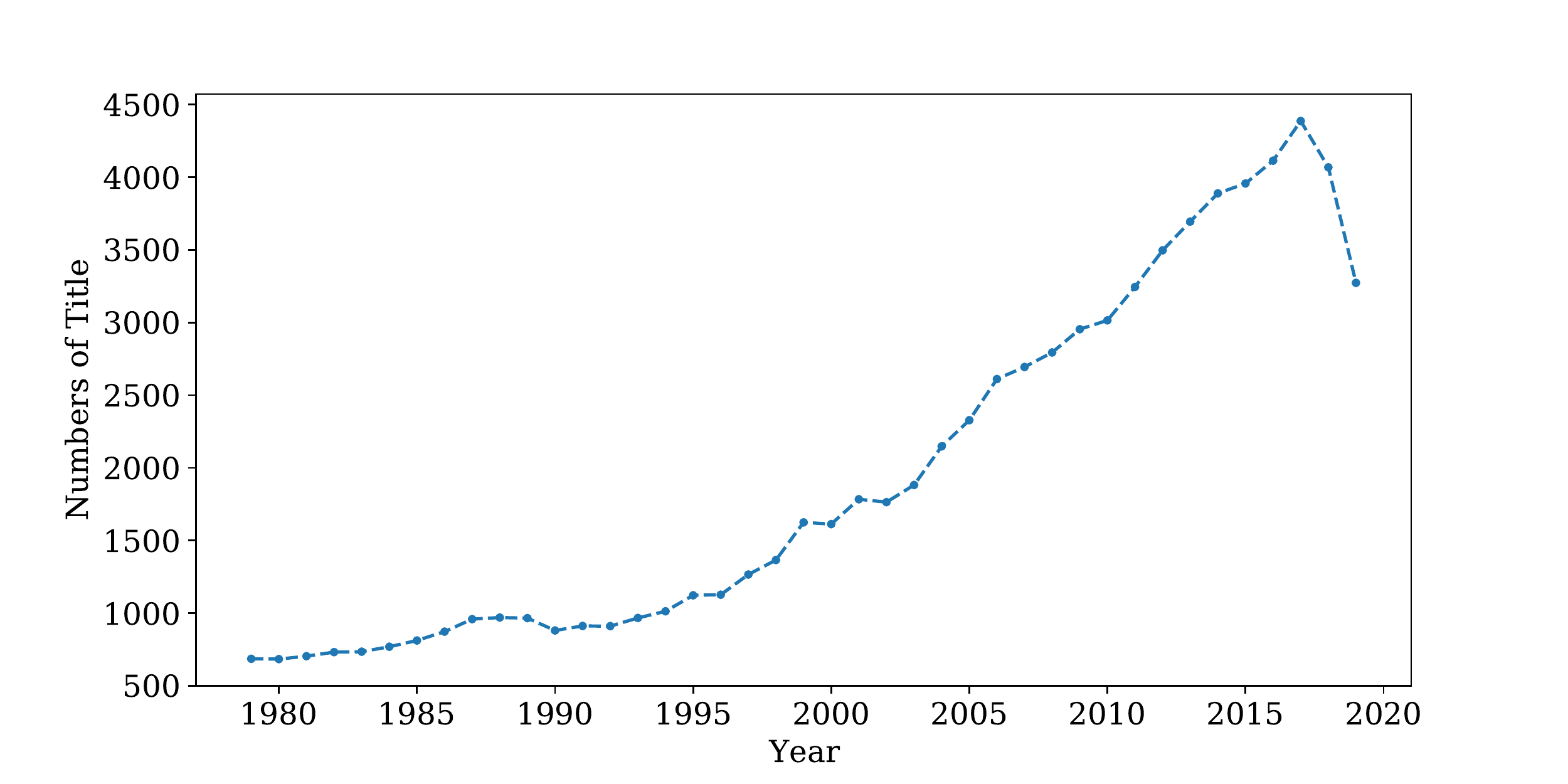}
		\caption{Number of movies with votes number larger then 100 votes, over time}
		\label{vt}
	\end{figure}
	
	\begin{figure}[H]
			\centering
		\includegraphics[width=0.7\textwidth]{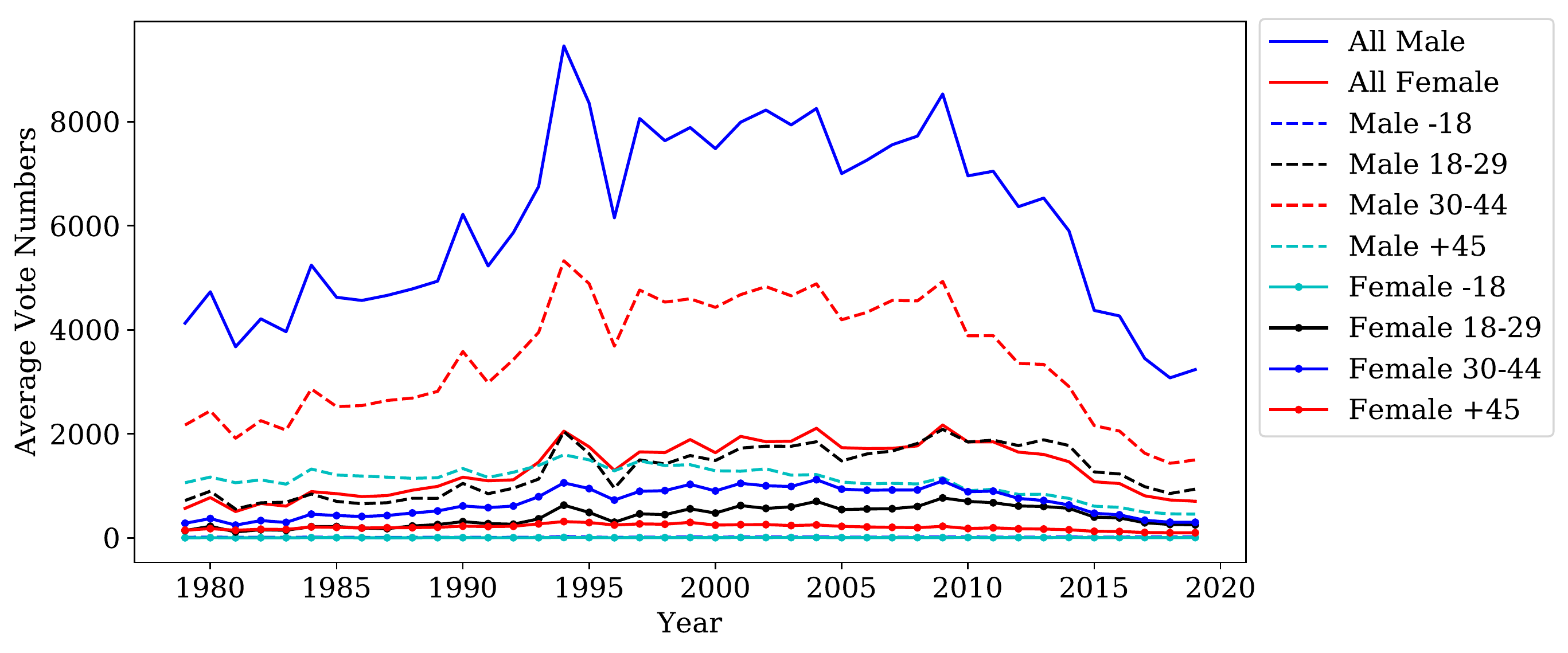}
		\caption{Averge of votes number of male and female voters in different age category }
		\label{f3}
	\end{figure}

	\begin{figure}[H]
			\centering
		\includegraphics[width=\textwidth]{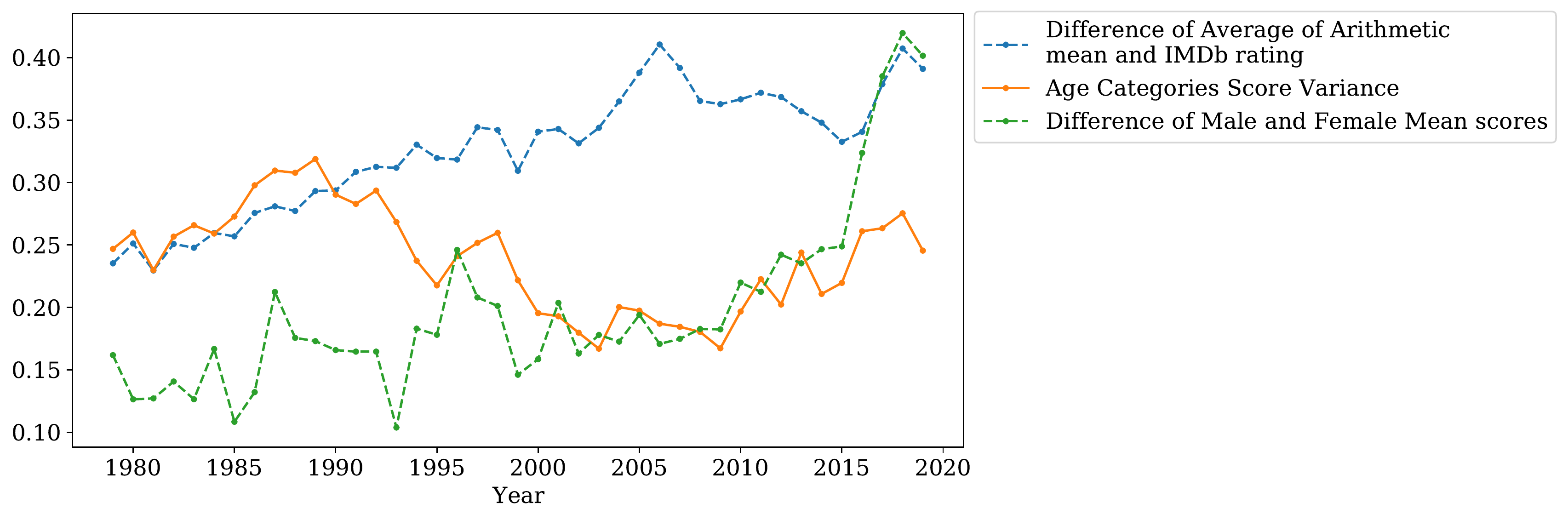}
		
		\caption{Trend of: difference of average of arithmetic mean and IMDb ratings, difference of male and female  mean scores, variance of age categories' score over time}
		\label{diff}
	\end{figure}

	\begin{figure}[H]
			\centering
		\includegraphics[width=\textwidth]{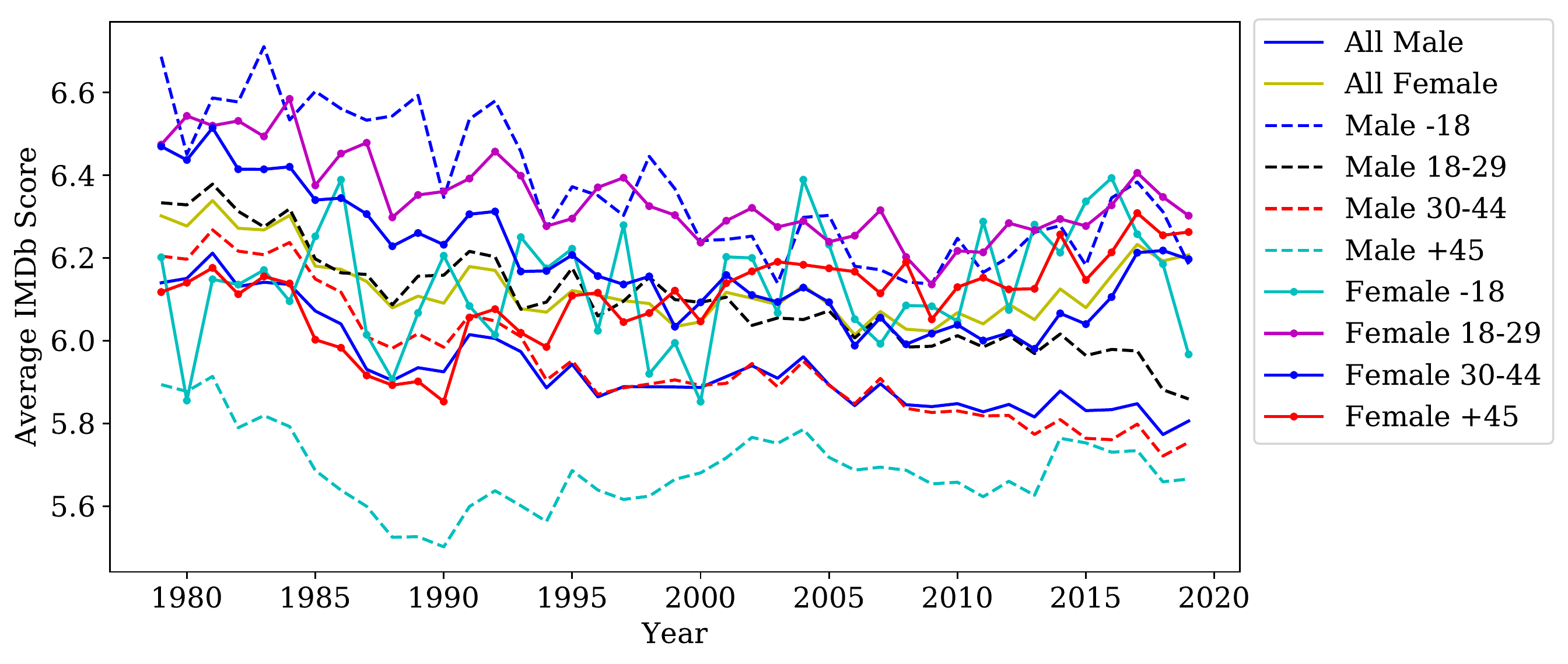}
		\label{sah}
		\caption{Average of scores given male and female voters in different age categories}
		\label{f3}
	\end{figure}
	\begin{figure}[H]
			\centering
		\includegraphics[width=0.8\textwidth]{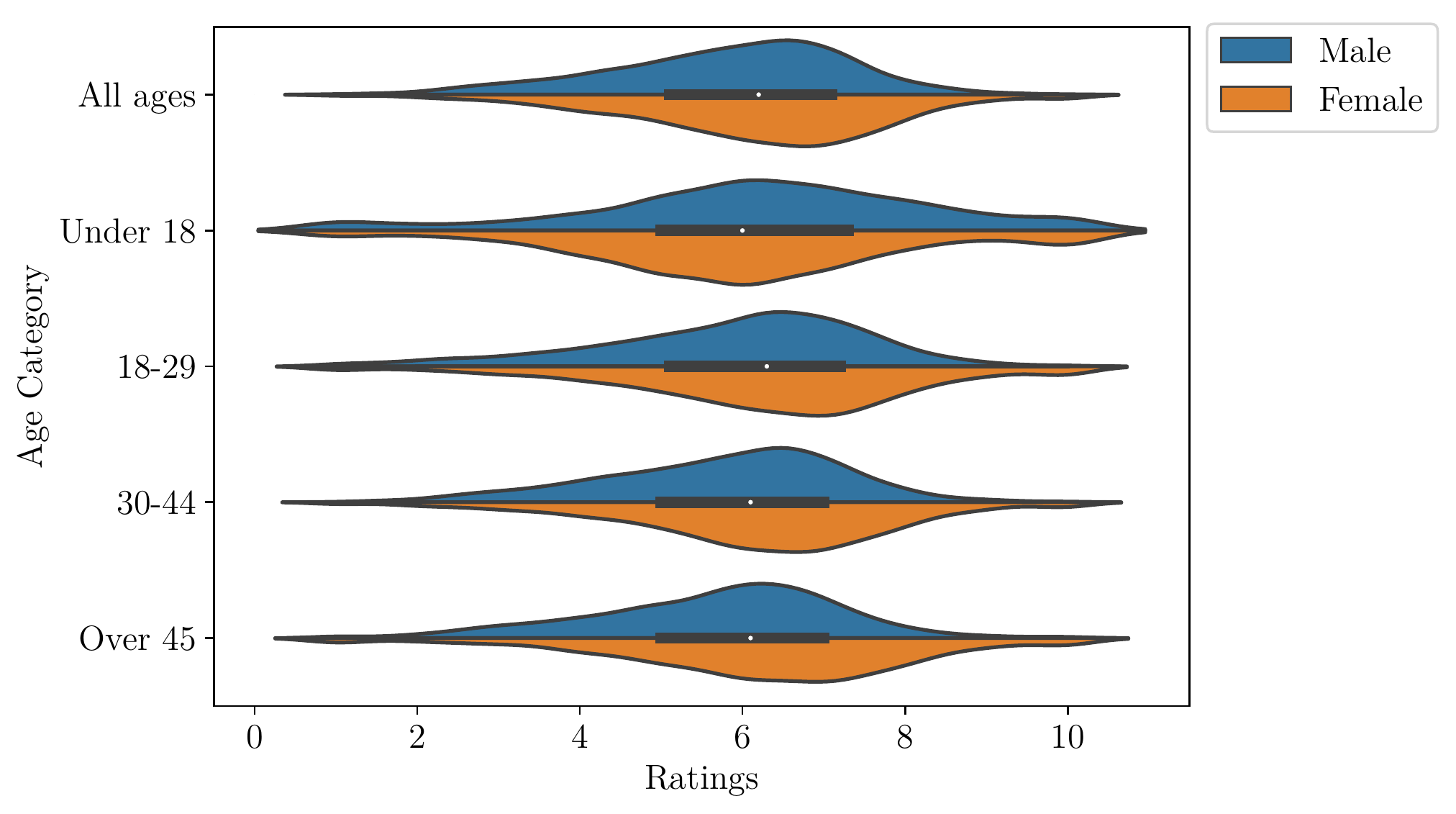}
		\caption{Quartiles and distributions of datings divided by age and gender}
		\label{vp}
	\end{figure}

	\begin{figure}[H]
			\centering
		\includegraphics[width=0.7\textwidth]{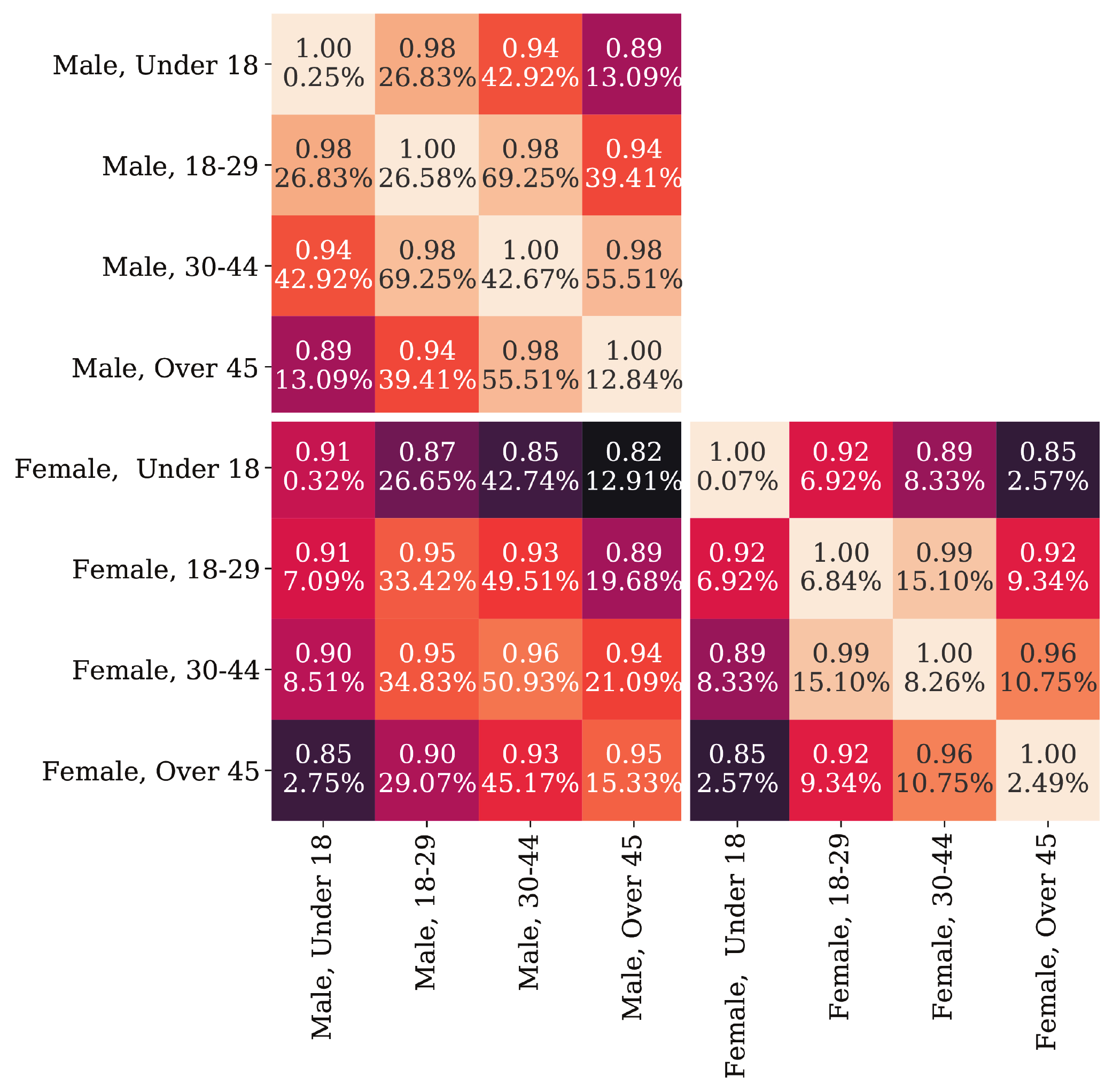}
		\caption{Correlations of scores given by males and females in different age category}
		\label{sh}
	\end{figure}

	\begin{figure}[H]
			\centering
		\includegraphics[width=0.8\textwidth]{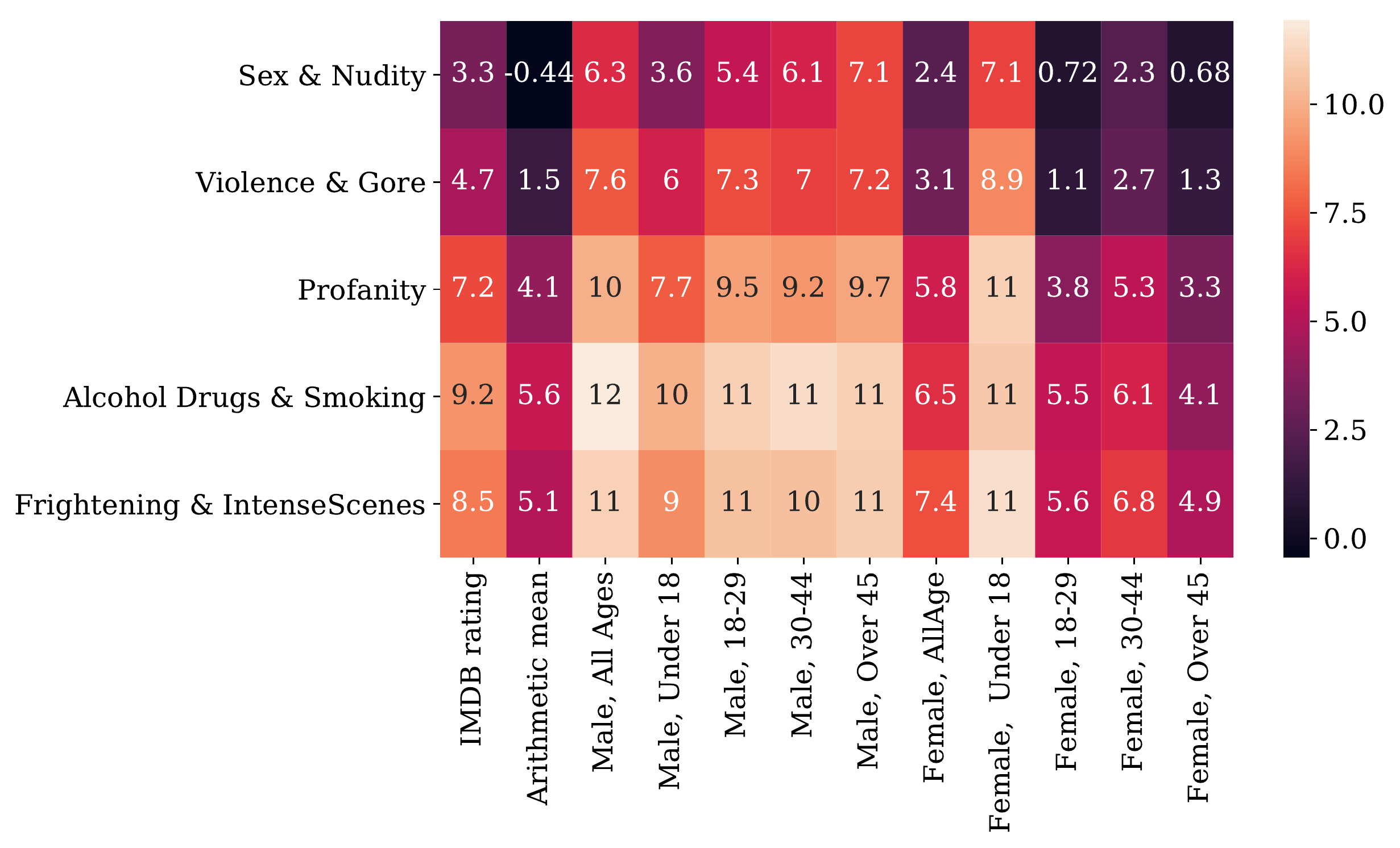}
		\caption{Correlations of parental guide items with IMDb ratings, Arithmetic mean and scores given by male and female in different age category - correlation values are multiplied by 100}
		\label{f3}
	\end{figure}

	\begin{figure}[H]
			\centering
		\includegraphics[width=0.5\textwidth]{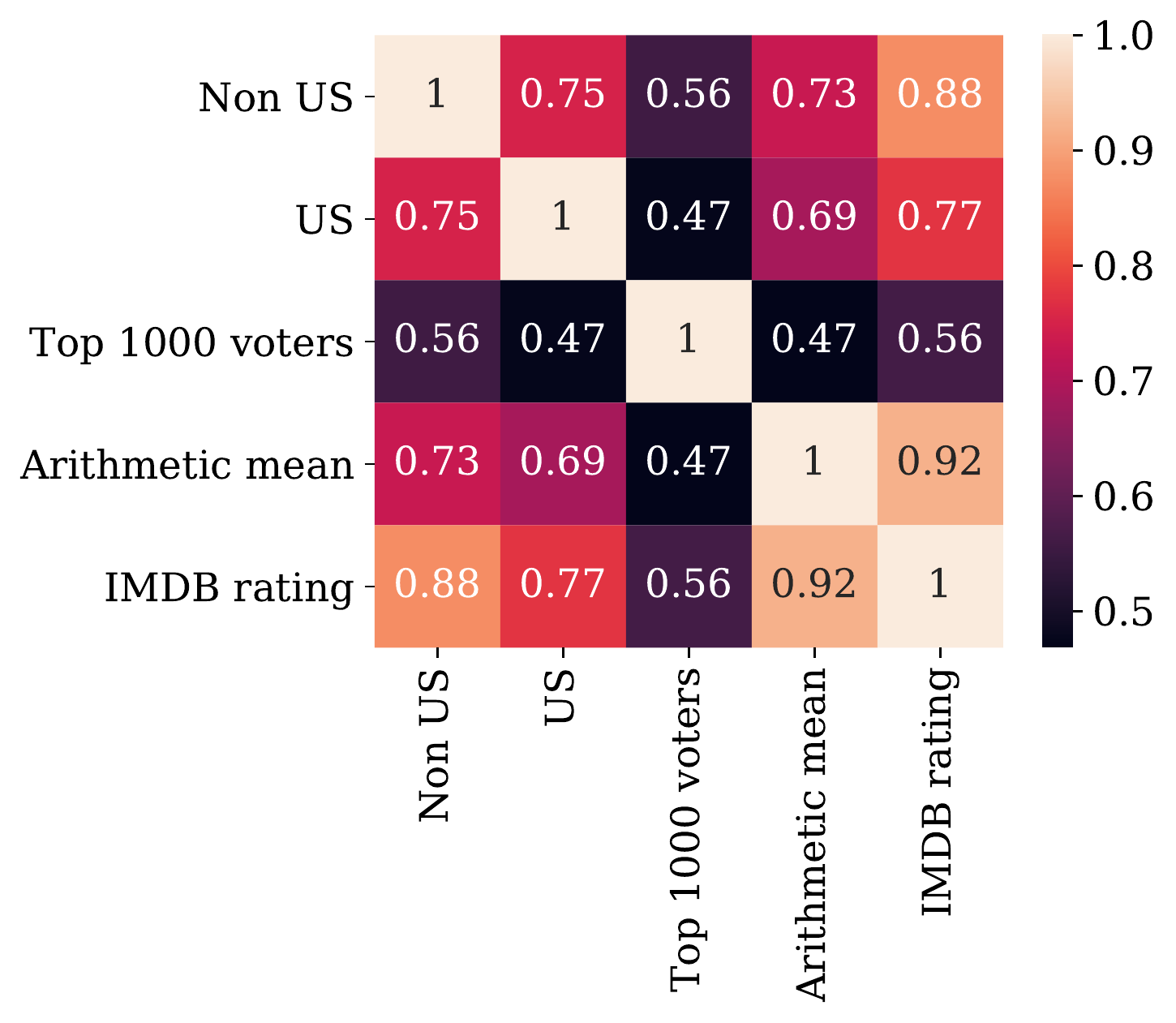}
		\caption{Correlations of  IMDb ratings, Arithmetic Mean and the Scores  US  and Non US and Top 1000 voters}
		\label{ff3}
	\end{figure}

	\begin{figure}[H]
			\centering
		\includegraphics[width=0.8\textwidth]{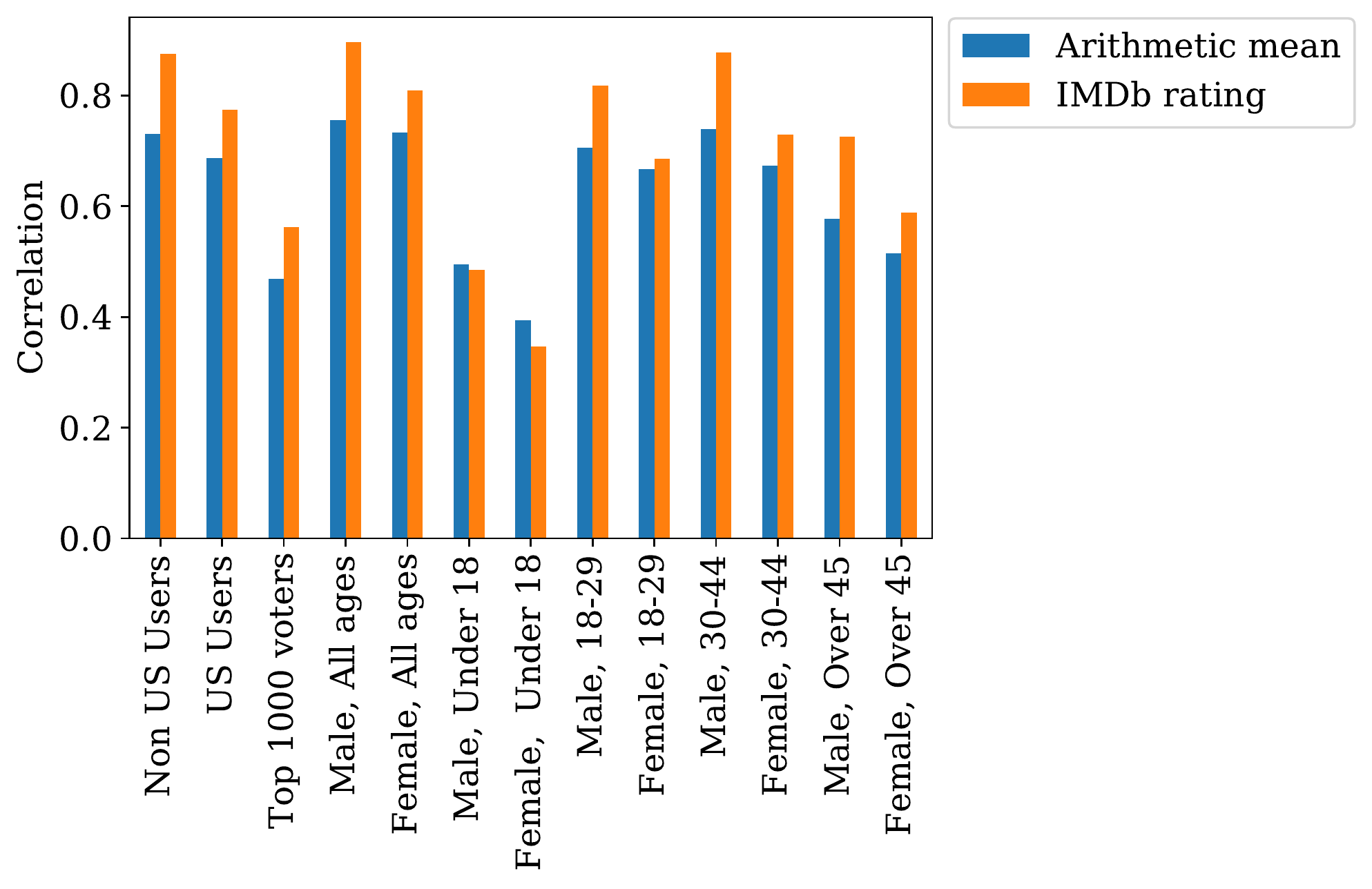}
		\caption{Correlations of US and Non-US and Top 1000 voters, Male and Females in Different age Category with IMDb ratings, Arithmetic Mean}
		\label{rating-AAM}
	\end{figure}
\end{center}
\end{widetext}
\end{document}